\documentclass[11pt]{article}%
\usepackage{geometry}
\usepackage{amsmath,amssymb,amsthm,amsfonts}
\usepackage{fancyhdr}
\usepackage{hyperref}
\usepackage{graphicx}
\usepackage{accents}
\usepackage[numbers]{natbib}
\usepackage{amsmath}
\usepackage{amsfonts}
\usepackage{amssymb}%
\providecommand{\U}[1]{\protect\rule{.1in}{.1in}}
\numberwithin{equation}{section}
\renewcommand{\footrulewidth}{0.4pt}

\begin{document}

\title{Systemic Risks in CCP Networks}
\author{Russell Barker, Andrew Dickinson, Alex Lipton, Rajeev Virmani\footnote{
All of the authors are employees of Bank of America. The views expressed in this paper are those of the authors and do not necessarily represent the views of Bank of America.}}
\date{April 1, 2016}
\maketitle

\abstract{We propose a model for the credit and liquidity risks faced by clearing members of Central Counterparty Clearing houses (CCPs).
This model aims to capture the features of: gap risk; feedback between clearing member default, market volatility and margining requirements;
the different risks faced by various types of market participant and the changes in margining requirements
a clearing member faces as the system evolves. By considering the entire network of CCPs and clearing members, we investigate the distribution of
losses to default fund contributions and contingent liquidity requirements for each clearing member; further, we identify wrong-way risks between
defaults of clearing members and market turbulence.}

\thispagestyle{fancy} 
\chead{}
\cfoot{}
\rfoot{Page:\ \thepage}
\renewcommand{\footrulewidth}{0.4pt}

\newpage

\section{Overview}

Since the Financial Crisis of 2007-2010 the
number of trades and the range of products that are cleared by CCPs has
increased enormously.

\begin{figure}[htbp]
	\centering
	\includegraphics[trim=2cm 2cm 4cm 2cm, clip=true, width=0.8\textwidth]{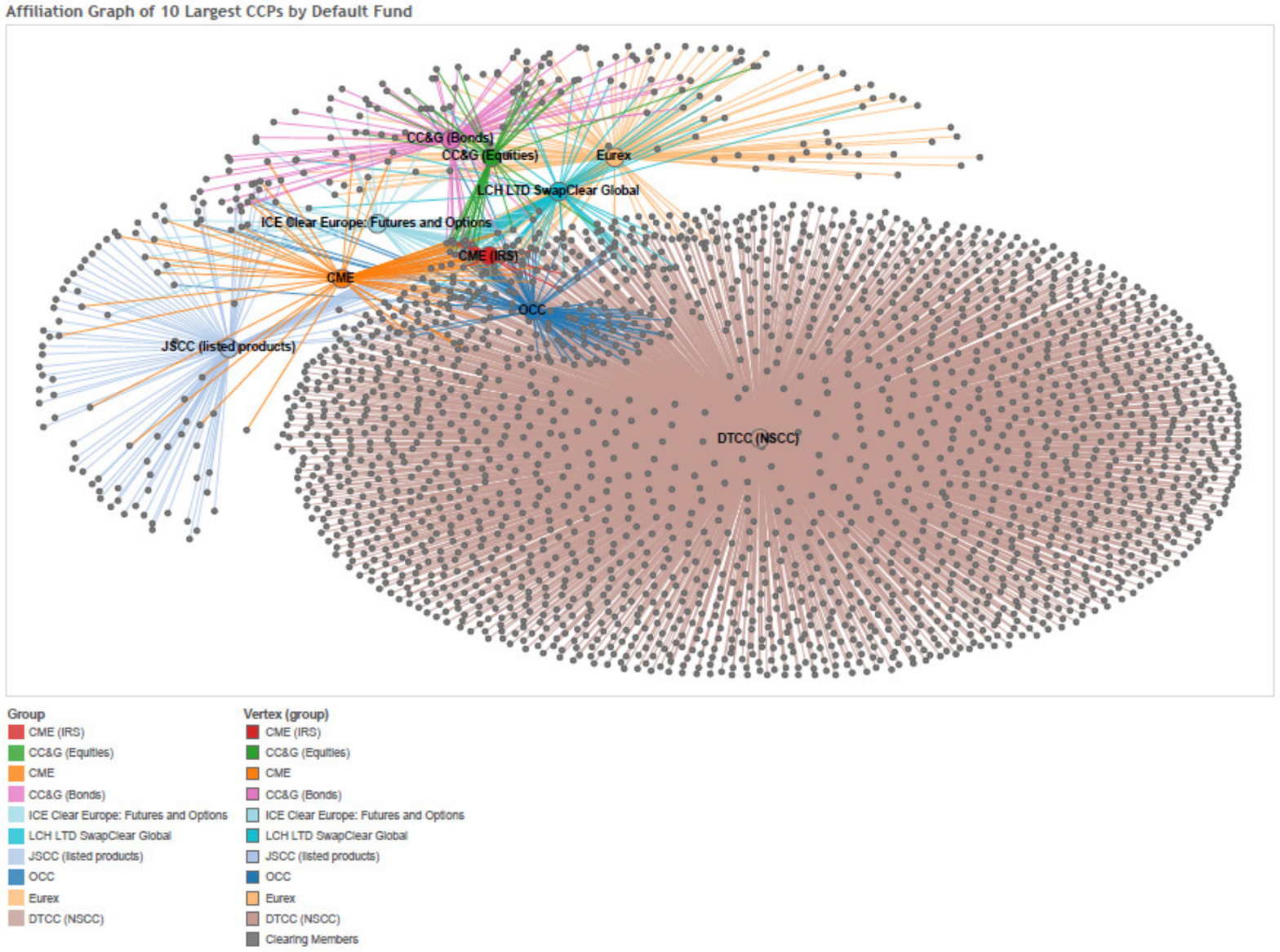}
	\caption{An Example Network of CCPs (coloured circles) and GCMs (black dots) with corresponding
	coloured edges denoting	membership}
	\label{fig:CCPNetwork}
\end{figure}

There is a clear need for a bank to assess
any potential impact of defaults of general clearing members (GCMs) through the
CCP network and, in particular, on themselves. However, understanding the
risk is a challenging one, since, it requires understanding the contingent
cash-flows between a large number of agents (hundreds of GCMs and
multiple CCPs), see Figure \ref{fig:CCPNetwork} for an example of a real-world CCP network.
Further, the inter-relationship between the GCMs with each
other via the CCPs is a complex one, requiring capturing the dynamic evolution 
of variation margin (VM), initial margin (IM), default fund contributions (DF), porting
of trades in the event of a member default and allocation of default losses. Further, the fact that a particular GCM's
CCP activity may only represent a fraction of their broader economic activity should be captured.
Although the system is too complex to analyze \emph{analytically}, it is viable to develop simulation models
which capture the contingent cash-flows between all agents
(including those related to margining and defaults), and address the following
important questions related to the broader application of central clearing to OTC
derivative portfolios:
\begin{itemize}
\item Potential systemic risks and contagion introduced by the interconnected nature of the system.
\item Liquidity issues driven by: P\&L; changes in margining; losses due to
default; CCP recapitalization.
\item The connection between market volatility and default likelihood.
\item Identification of the key points of failure.
\item The magnitude of scenarios proving
sufficiently large in order for a given clearing member to incur a
loss or suffer liquidity issues.
\end{itemize}

One of the novel points of the model proposed in this paper is that
it considers the entire network of CCPs and GCMs which, given its
size and complexity, is somewhat challenging and yields some important insights:
\begin{itemize}
\item There are material cross-risks between the default of GCMs and market volatility
which must be captured in order to realistically assess default losses and 
contingent liquidity requirements.
\item Our results do not support the fear that the move from bilateral clearing
to central clearing of OTC derivatives poses a significant threat of contagion through the central
counterparties, primarily attributable to the magnitude of risks being a comparatively small
proportion of the capital held by the diversified financial institutions dominating the CCP membership.
\end{itemize}

A wide variety of models have been developed in order to quantify potential
exposure of CCPs. These models can be divided into three main categories:
statistical, optimization and option pricing based models.

Statistical models typically assume simple underlying dynamics, such as
Geometric Brownian Motion (GBM) and derive the probability for
the Initial Margin (IM) to be exceeded within a given time horizon. A typical
paper along these lines is Figlewski (1984) \cite{Figlewski} who calculated
the probability of a margin call given a certain percentage of initial and
maintenance margin.

Optimization models, as their name suggests, try to set margins in a way that
achieves an appropriate balance between resilience of CCPs and costs to their
members. In particular, Fenn and Kupiec (1993) \cite{FK} and Baer, France and
Moser (1995)\cite{BFM} built models for minimizing the total sum of margin,
settlement and failure costs.

Option pricing based models capitalize on the fact that the exposure profile
of a CCP is approximately equivalent to the payoff of a strangle, representing
a combination of a call and a put option. A GCM has a theoretical opportunity
to default strategically if the contract loses more value than the posted IM.
Day and Lewis (1999)\cite{DL} considered margined futures positions as barrier
options and estimated prudent margin levels for oil futures for the period
1986 to 1991.

Since the most important problem from the CCP standpoint is to analyze losses
conditional on exceeding margin, Extreme Value Theory (EVT) is well suited for
this purpose and had been exploited by several researchers, see, e.g.,
Broussard (2001)\cite{Broussard} and Longin (1999)\cite{Longin}. Whilst it is
relatively easy to use EVT to set up margins for a single contract, it is much
more difficult at a portfolio level, hence CCPs tend not to use EVT
directly. Accordingly, in many cases CCPs use Standard Portfolio Analysis of
Risk (SPAN) methodology instead, see Kupiec (1994)\cite{Kupiec}. While
intuitive and appealing, SPAN has significant limitations when used to
calculate margins for complex portfolios. Accordingly, value at risk (VaR)
based margining system gained considerable popularity. Application of VaR
methods to large complex portfolios was discussed by Barone-Adesi et al.
(2002)\cite{Barone}.

Duffie and Zhu (2011) \cite{Zhu} discuss the premise that central clearing of derivatives can substantially reduce counterparty risk. They argue that some of these benefits are lost through a fragmentation of clearing services due to no allowance for ‘inter-operability’ across asset classes and/or CCPs.  They measure the tradeoff between two types of netting opportunities: bilateral netting between counterparties across assets, versus multilateral netting among many clearing participants across a single class of derivatives, and argue that benefit of one over the other depend on the specifics of the clearing process.

In Glasserman, Moallemi and Yuan (2014), the authors discuss issues pertinent to managing systemic risks in markets cleared by multiple CCPs.
Since each CCP charges margins based on positions of a clearing member, it creates incentives for swaps dealers to split their
positions among multiple CCPs. This splitting causes the `hiding' of potential liquidation costs from each individual CCP, thus
underestimating these costs.

Borovkova and El Mouttalibi (2013) \cite{Borovkova} analyze systemic risk in CCPs by utilizing a network approach. They show that the effect of CCPs on the stability of the financial system is rather subtle. According to the authors, stricter capital requirements have a clearer and stronger positive impact on the system than mandatory clearing through CCPs.

Cont and Avellaneda (2013) \cite{Cont} develop an optimal liquidation strategy for disposing a defaulted member’s portfolio. Parts of the portfolio are sold by auction, parts are unwound, and the rest is sold on the market. Their approach evaluates issues of market liquidity under adverse conditions. It models an auction with constraints applied on how many positions can be liquidated on a given day, and detemines an optimal sale strategy to minimise market risk. The authors construct an objective function and optimise over SPAN scenarios using linear-programming.

Cumming and Noss (2013)\cite{Cumming} assess the adequacy of CCPs' default
resources from the perspective of the Bank of England. They argue that the best way
to model a CCP's exposure to a single GCM in excess of its IM is by applying
Extreme Value Theory (EVT). The authors propose using a simple analogy between
the risk faced by a CCP's default fund and the one borne by a mezzanine tranche
of a Collateralized Debt Obligation (CDO). The authors use an established
framework to model co-dependence of defaults based on a gamma
distribution; although the reader should note the caveat that 
the model assumes that exposures and defaults are uncorrelated, which is
unlikely to be the case. Moreover, the analogy between CCPs and CDOs is a loose one,
GCMs of a CCP are very different from credits used to construct CDOs. Still,
the model is unquestionably a useful step towards building a proper top-down
statistical framework for assessing of the risk of a CCP's member exposures.

Murphy and Nahai-Williamson (2014) \cite{PRACCP} discuss approaches to the
analysis of CCP DF adequacy. They start with a range of design choices for the
default waterfall and discuss regulatory constraints imposed on the waterfall.
The authors concentrate on the cover 2 requirement because it is a minimum
internationally acceptable standard for a CCP. Their contribution is two-fold
- (a) it is shown how to use market data to estimate the complete distribution
of a CCP's stressed credit risk; (b) the prudence of cover 2 as a function of
the number of GCMs is examined.

Elouerkhaoui (2015) \cite{Elouerkhaoui} develops a method for calculating
Credit Value Adjustment (CVA) for CCPs using a CDO pricing approach by defining the payoff of the CCP's waterfall and using the Marshall-Olkin correlation model to compute it. The author considers a CCP as one of the counterparties a GCM faces
in the market place and derives the Master Pricing Equation with bilateral
CVA, as well as Funding Value Adjustment (FVA) and Margin Value Adjustment
(MVA). Cr\'{e}pey (2015) \cite{Crepey} pursues a similar approach. While well thought through, their approach suffers from the fact that defaults of GCMs (and hence the CCP
itself) are not directly linked to the behavior of the cleared product.

We feel that although a lot of advances have been made in the recent
literature some of the most important features of the CCP universe have been
missed. The first is the feedback mechanism that intrinsically links GCM
default, market turbulence, and liquidity calls on market participants. The
second is the individual nature of different clearing members: from large
diverse financial institutions where markets will make up a minority of their
business to proprietary funds for which a default event will be purely driven
by margin calls on cleared trades. The third is the interconnectedness of the
CCPs themselves meaning that it is important to model the network in its
entirety. Finally it is important to model the changes in IM\ and DF
requirement as the system evolves; this is particularly important for
modelling liquidity considerations.

The approach we follow is to use the minimum amount of information necessary to
analyse the risk of contagion or a liquidity crunch in the CCP framework but
still build a realistic simulation of what might occur in a stressed
situation. For each GCM and each CCP we need to model the loss over IM and DF
if that GCM were to default in a specific market scenario and how to
distribute that loss to other GCMs. We also need to be able to model the
circumstances of a GCM default given a market scenario and link this to the
reduction in the GCM's capital due to losses on a number of CCPs.

\begin{figure}[htbp]
	\centering
	\includegraphics[trim=0.5cm 0.5cm 0.5cm 0.5cm, clip=true, width=0.7\textwidth]{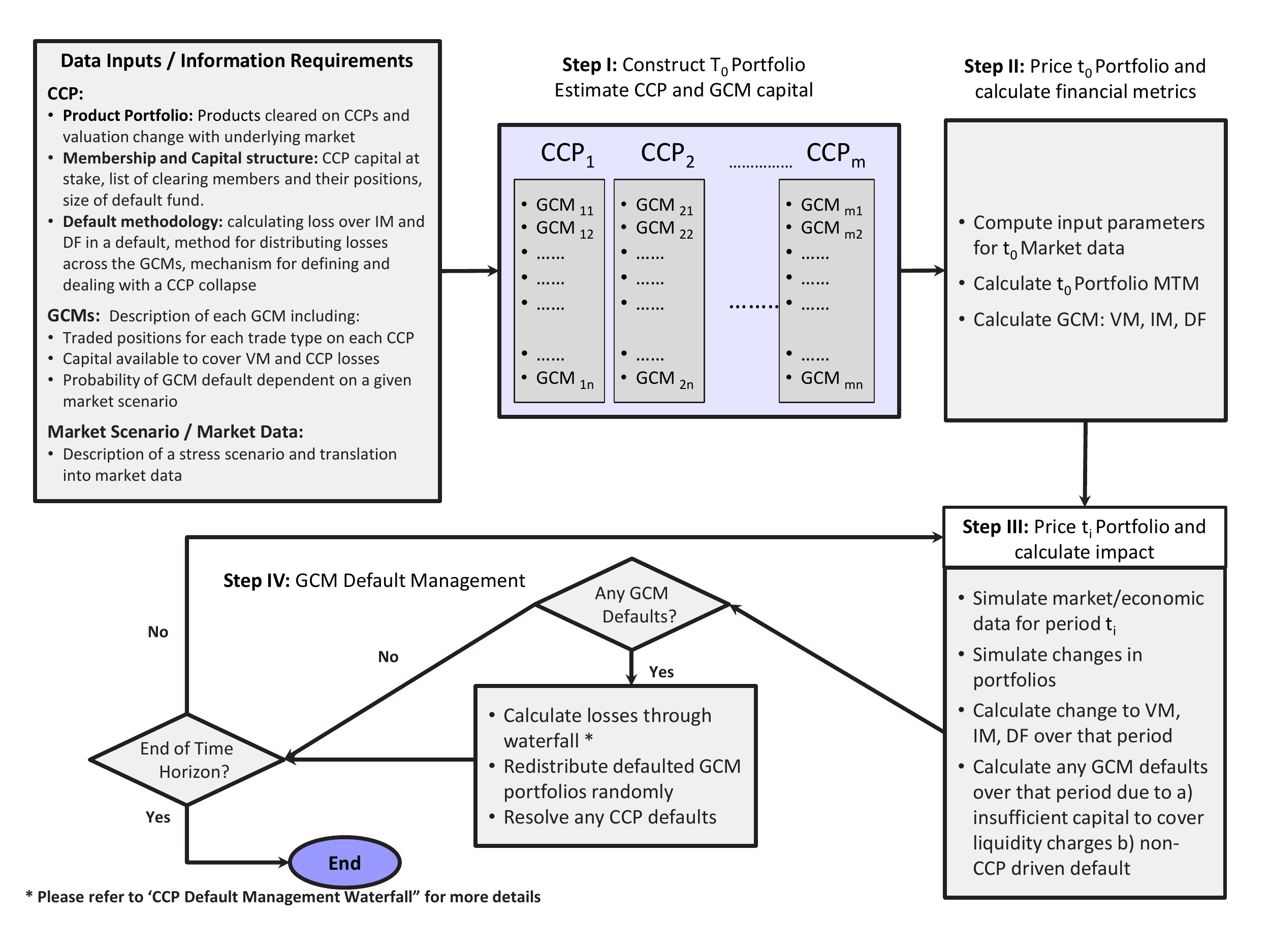}
	\caption{Schematic of CCP Model}
	\label{fig:modelScehmatic}
\end{figure}

In this paper, we develop a simulation framework to investigate the risks
associated to central clearing. The paper is
structured as follows: in Section \ref{sec:margCalc}, we discuss
margining and its modelling; in Section \ref{sec:GCMPort}, we present the process
by which we generate the portfolios of clearing members given the partial information
a particular GCM possesses; in Section \ref{sec:ScenarioGeneration}, we present 
the simulation of the underlying market variables and the feedback mechanism
used to generate realistic co-dependence between volatilities and defaults;
we present numerical results in Section \ref{sec:results} and conclude with
Section \ref{sec:conclusions}.

\section{Margin Calculation}\label{sec:margCalc}

CCPs set-up extensive processes to manage the default of any GCM including
requiring their GCMs to post an Initial Margin (IM) and a Default Fund (DF)
contribution as well as Variation Margin (VM) to cover the MtM moves of the
exposures together with a Risk Waterfall process that stipulates how any eventual losses
would be distributed among the defaulting clearing member, the non-defaulting
members and the CCP itself. \ Given a set of market data and a portfolio of
trades we need to be able to calculate VM, IM and the DF for the total set of
GCM portfolios on a given CCP.

We represent the state of the market at time $t$ by $\mathbf{X}(t)=\left(X_{1}\left(t\right),\dots,X_{n}\left(t\right)\right)^{T}$,
where $X_{i}(t)$ represents a financial quantity such as a par swap, spot FX rate,
credit spread etc. We describe the generation of these scenarios in Section \ref{sec:ScenarioGeneration}.

The incremental VM called over the time interval $\left[t_{i}, t_{i+1}\right]$ for the portfolio held by $GCM_{k}$, with
$CCP_{j}$, is given by the change in mark-to-market

\[
VM_{CCP_{j}}^{GCM_{k}}\left(t_{i+1}\right) - VM_{CCP_{j}}^{GCM_{k}}\left(t_{i}\right)=%
{\displaystyle\sum\limits_{\phi\in\Phi^{GCM_{k}}_{CCP_{j}}\left(t_{i} \right) } }
V_{\phi} \left(\mathbf{X}\left(t_{i+1}\right),t_{i+1}\right) - V_{\phi} \left(\mathbf{X}\left(t_{i}\right),t_{i}\right)
\]
where the summation is over all trades, $\phi$, in the portfolio $GCM_{k}$ holds with $CCP_{j}$ at time $t_{i}$ and
$V_{\phi}\left(\mathbf{X}\left(t\right),t\right)$ is the value of trade $\phi$ at time $t$ in market state $\mathbf{X}\left(t\right)$.

We need to construct a fast method for calculating $GCM_{k}$'s $IM$ on
$CCP_{j}$ based on its portfolio and the market at time $t_{i}$. Generally,
the IM is dominated by a VaR/CVaR component of the portfolio over a
set of market changes derived from a historic time period supplemented by a set
of deterministic add-ons for liquidity, basis risk etc. c.f. Avellaneda and
Cont \cite{Cont}. We split the IM into a VaR/CVaR component and a set of
add-ons. In practice, the VaR/CVaR component is calculated across the portfolio losses as follows:

\begin{itemize}
\item {\small Create a set of scenarios by looking at a set of 5-day change in
market data over some historic period.}

\item {\small Weight these changes by a multiple of the ratio of current to
historic realized volatility.}

\item {\small Create new scenarios using the current market data and the set
of market changes and calculate the 5-day loss on the portfolio for each
scenario.}

\item {\small Calculate the VaR or CVaR using these.}
\end{itemize}

We calculate separately the VaR/CVaR component using regression against a collection
of representative portfolios and then
apply the add-ons deterministically: firstly, we evaluate the IM
on a set of portfolios which are sufficiently small so as to not incur any add-ons
and the IM on these
are calculated using the full IM calculation process; secondly the IM for the current portfolio is
calculated as
\begin{align}
IM_{CCP_{j}}^{GCM_{k}}\left(t\right)
= VaR\left(\left\{\mathbf{X}\left(u\right)\right\}_{u\leq t};\mathbf{a}\left(t\right)\right) + AddOn\left(t\right)
\label{eq:IM}
\end{align}
where $\mathbf{a}\left(t\right) = \left(a_{1}\left(t\right),\dots,a_{n}\left(t\right)\right)^{T}$
represents the regression coefficients of the portfolio held at
time $t$ against the small benchmark portfolios for which IM has been calculated in the first step,
$VaR\left(\left\{\mathbf{X}\left(u\right)\right\}_{u\leq t};\mathbf{a}\left(t\right)\right)$ represents the VaR component of the IM
for a portfolio represented by these regression coefficients and
adjusted for new simulated market data taking into account:
\begin{enumerate}
\item The change in the ratio of current market to historic volatilities.
This can be estimated by keeping track of the historic multiplier and
adjusting appropriately.
\item\label{en:update} If the new scenario creates a loss large enough to replace one of the
VaR or CVaR elements\footnote{We assume that a VaR/CVaR scenario does not roll
off the historic period specified; this is a reasonable assumption as most
CCPs include the 2007-2012 period and have ensured that this will not drop out
of the historic period in the near future and the time horizons for our simulations are
around 1 year.}.
\end{enumerate}
We ignore the contribution to the volatility of the add-on in (\ref{eq:IM}) and we freeze it to the time 0 value.

To take Point \ref{en:update} above into account, we heuristically model the historical loss distribution
of the portfolio by a one-parameter family of distributions (such as the family of centred Gaussian distributions)
fitted to the level of the IM at the previous time-step and update according
to whether the realization of the loss over the current time step is sufficiently extreme.

The total default fund at time $t$ is given by a `Cover 2' principle
\begin{align}
DF_{CCP_{j}}\left(t\right) = \max_{\sigma \in S_{j}} \max_{k \neq l} \left[LOIM_{CCP_{j}}^{GCM_{k}}\left(t,\sigma\right) + LOIM_{CCP_{j}}^{GCM_{l}}\left(t,\sigma\right) - K_{CCP_{j}}\right]^{+}
\end{align}
where the maximum is over all stress scenarios, $S_{j}$, for $CCP_{j}$ and
distinct pairs of (surviving) GCMs $GCM_{k},GCM_{l}$ and the loss over IM, 
$LOIM_{CCP_{j}}^{GCM_{k}}\left(t,\sigma\right)$, is given by
\begin{align*}
LOIM_{CCP_{j}}^{GCM_{k}}\left(t,\sigma\right) = \left[\sum_{\phi \in \Phi_{CCP_{j}}^{GCM_{k}}\left(t\right)} \left(V_{\phi}\left(t,\mathbf{X}\left(t\right)\right)
- V_{\phi}\left(t,\mathbf{X}^{\sigma}\left(t\right)\right)\right) + IM_{CCP_{j}}^{GCM_{k}}\left(t\right)\right]^{+}
\end{align*}
the summation is over trades, $\phi$ in the portfolio $\Phi_{CCP_{j}}^{GCM_{k}}\left(t\right)$ and
$V_{\phi}\left(t,\mathbf{X}^{\sigma}\left(t\right)\right),V_{\phi}\left(t,\mathbf{X}\left(t\right)\right)$
are the values of $\phi$ at time $t$ in market state $\mathbf{X}\left(t\right)$ with and without
the stress scenario, $\sigma$, applied. We remark that although CCP's define a large
number of stress scenarios, typically, there are comparatively few `binding' scenarios
and the scenarios may be replaced with a significantly smaller subset.

Although there may be a variety of ways to allocate the DF amongst members, we shall assume
it to be pro-rated among the surviving members according to their IM:
\begin{align*}
DF_{CCP_{j}}^{GCM_{k}}\left(t\right) =& \frac{\chi_{\tau_{k} > t} IM^{GCM_{k}}_{CCP_{j}}\left(t\right)}
{\sum_{l} \chi_{\tau_{l} > t} IM^{GCM_{l}}_{CCP_{j}} \left(t\right)}
DF_{CCP_{j}}\left(t\right)
\end{align*}
where $\chi_{\tau_{k} > t}$ denotes the survival indicator of $GCM_{k}$.

\section{GCM Portfolios}\label{sec:GCMPort}
In this section, we take the perspective of a particular (but arbitrary)
clearing member and discuss how it may assess its risks making use of the
partial information available to it. It is not uncommon for a banking group
to have multiple subsidiaries, each of which are distinct clearing members. We write $\left\{ GCM_{k}\right\} _{k\in K}$ for these subsidiaries whose
positions with each CCP are known to the group. To aid discussion let us
introduce a name for this banking group, `XYZ Bank'. We refer to the \emph{
gross} notional as the total notional over long and short positions and the 
\emph{net} notional to be the difference. The reader should note that risks
are ultimately determined by net positions and hence net notionals are of
primary concern; however, gross notionals provide useful and important
information of accumulated historical volumes.

Although XYZ is potentially exposed to the positions of other members in the
event of their default, the positions of other members are unknown. However,
gross notionals for certain categories of derivatives, aggregated across all
members, are published by the CCP where the categories are typically
discriminated by the type of trade, currency and tenor. For example, a CCP
may publish the aggregate gross notional for fixed vs 6M EURIBOR swaps for
tenors in the range 2y to 5y (alongside other aggregates). We wish to make
use of these aggregate gross notionals as a measure of the relative scale of
the exposures of the CCP in different trade types, currencies and tenors.
For this reason, it is appropriate to align our methodology to the
categorization used by the CCP's when reporting aggregate gross notionals.
Below, we will fix a particular category, $\pi \in \Pi$ where $\Pi$
represents the set of all categories used by the CCP to disclose aggregate
gross notionals.

To assess XYZ's risks, we propose a randomization scheme to explore the
space of valid configurations of the unknown positions of other GCM's,
subject to the constraints of reproducing the known information: values
related to XYZ's positions and the aggregate gross notionals published by
the CCPs. To each GCM, $k$, we assign a rank $J_{k}\in \left\{ 1,\dots
,n\right\} $, based upon data sourced from publicly available information,
such as financial statements. We then fit a two-parameter exponential
distribution to the gross notional of the members, motivated by the analysis
of Murphy and Nahai-Williamson \cite{PRACCP} 
\begin{align}
\sum_{k=1}^{n}\beta ^{\pi }\exp \left( -\alpha ^{\pi }J_{k}\right) =& N^{\pi
}  \label{eq:expFit1} \\
\sum_{k\in K}\beta ^{\pi }\exp \left( -\alpha ^{\pi }J_{k}\right) =&
N_{K}^{\pi }  \label{eq:expFit2}
\end{align}%
where $K$ is the set of indices of XYZ's members, $N^{\pi },N_{K}^{\pi }$
are the gross notional for category $\pi $ aggregated over all members and
XYZ's members, respectively. The system is solved numerically\footnote{%
Abel-Ruffini suggests that there is no analytic solution for this set of
equations except in a few trivial cases.} for $\alpha ^{\pi },\beta ^{\pi }$
and we abbreviate the fitted net notional for $k$ as 
\[
N_{k}^{\pi }=\beta ^{\pi }\exp \left( -\alpha ^{\pi }J_{k}\right) .
\]
We generate randomized net notionals, $\left\{ \Delta _{k}^{\pi
}\right\} _{k=1}^{n}$, such that long and short positions net
\begin{align}
\sum_{k=1}^{n}\Delta_{k}^{\pi }&=0 \label{eq:parity}
\end{align}
and
\begin{align}
\Delta_{k}^{\pi } &\in \left[-R N_{k}^{\pi}, R N_{k}^{\pi} \right], \quad \forall k \label{eq:deltaLimit} \\
\Delta_{k}^{\pi} &= \delta_{k}^{\pi}, \quad \forall k \in K \label{eq:bacDelta}
\end{align}
where $R \in \left[0,1\right]$ is a parameter controlling the relative size of the net and gross positions
and $\left\{\delta _{k}\right\}_{k\in K}$, are the known net positions for XYZ. $R$ may be thought of as a
proportional trading delta limit and, for reasons of parsimony, we assume that
the parameter does not depend upon the GCM, $k$, nor the product
classification, $\pi$. Of course, this assumption could be weakened if
appropriate.

Introduce the negative total sum of all known positions, $\bar{\Delta}^{\pi
}=-\sum_{k\in K}\delta _{k}^{\pi }$, define the ratio $\ r^{\pi}=\bar{\Delta}^{\pi
}/\sum_{k\notin K}N_{k}^{\pi }$, and use it to proportionally allocate $\bar{\Delta}^{\pi
}$ among GCMs with $k\notin K$, $\ \bar{\Delta}_{k}^{\pi }=r^{\pi}N_{k}^{\pi }$.
Without loss of generality we assume that $\left\vert r^{\pi}\right\vert <R$. For
each $k\notin K$ consider the interval $I_{k}^{\pi}=\left[ -\left(R-\left\vert
r^{\pi}\right\vert\right) ,\left(R-\left\vert r^{\pi}\right\vert\right) \right] $ and generate 
independent random
numbers $u_{k}^{\pi}$ uniformly distributed on $I_{k}^{\pi}$. Define the following
quantities%
\begin{equation}
U^{\pi}=\sum_{k\notin K}u_{k}^{\pi}N_{k}^{\pi},\ \ \ V^{\pi}=\sum_{k\notin K}u_{k}^{\pi}N_{k}^{\pi}\chi%
_{u_{k}^{\pi}U>0},\ \ \ W^{\pi}=\frac{U^{\pi}}{V^{\pi}},  \label{Eq2}
\end{equation}%
where $\chi$ is the indicator function. Since $u_{k}^{\pi}$ possesses a density, it is clear that $V^{\pi}\neq 0$
almost surely and $W^{\pi}$ is
well-defined, and $0<W^{\pi}\leq 1$. Define net position of the $k$-th GCM as
follows
\begin{equation}
\Delta _{k}^{\pi }=\ \bar{\Delta}_{k}^{\pi }+u_{k}^{\pi}\left( 1-W^{\pi}\chi%
_{u_{k}^{\pi}U^{\pi}>0}\right)N_{k}^{\pi} =(r^{\pi}+u_{k}^{\pi}\left( 1-W^{\pi}\chi%
_{u_{k}^{\pi}U^{\pi}>0}\right))N_{k}^{\pi } .  \label{Eq3}
\end{equation}%
In words, we proportionally reduce positions for GCMs with $u_{k}^{\pi}$ having
the same sign as $U^{\pi}$, and keep positions for other GCMs fixed. $\allowbreak $%
A simple calculation yields%
\begin{equation}
\sum_{k\notin K}\Delta _{k}^{\pi }=\bar{\Delta}^{\pi}+U^{\pi}-%
\frac{U^{\pi}}{V^{\pi}}V^{\pi}=\bar{\Delta}^{\pi},  \label{Eq4}
\end{equation}%
\begin{equation}
\begin{array}{c}
\left\vert \Delta _{k}^{\pi }\right\vert =\left\vert \Delta _{k}^{\pi }-\ 
\bar{\Delta}_{k}^{\pi }+\ \bar{\Delta}_{k}^{\pi }\right\vert \leq \left\vert
\Delta _{k}^{\pi }-\ \bar{\Delta}_{k}^{\pi }\right\vert +\left\vert \ \bar{%
\Delta}_{k}^{\pi }\right\vert  \\ 
\leq \left\vert u_{k}\right\vert +\left\vert r^{\pi}\right\vert N_{k}^{\pi }\leq
\left( R-\left\vert r^{\pi}\right\vert +\left\vert r^{\pi}\right\vert \right) N_{k}^{\pi
}=RN_{k}^{\pi },%
\end{array}
\label{Eq5}
\end{equation}%
so that both conditions for $\Delta _{k}^{\pi },\ k\notin K$, are satisfied.

\section{Scenario Generation}\label{sec:ScenarioGeneration}

In this section, we describe the model for the underlying market variables that is used to evolve
the system of GCMs given the initial positions generated according to the scheme presented in Section
\ref{sec:GCMPort}. We wish to ensure that the model is sufficiently rich so as to:
\begin{enumerate}
\item Support jumps, allowing for comparatively large changes on short time scales, including
systemic jumps affecting all market variables simultaneously,
\item Reflect that periods of high default rates will be accompanied by high market volatility (as was observed during
the crisis).
\end{enumerate}
For these reasons, we will propose a regime-switching model with regimes driven by the number
and size of realized defaults.

\subsection{Regime-Dependent Drivers}
For each GCM, $GCM_{k}$, we introduce a weight $w_{k}>0$ representing the financial significance 
of the GCM to the others and normalized so that $\sum_{k} w_{k} = 1$. Practically, we set these weights
to be proportional to the balance sheet assets of each GCM.

We introduce a stress indicator, $\Xi_{t}$, by
\begin{align*}
\Xi_{t} = \sum_{k} w_{k} e^{-\theta_{\Xi} \left\{t - \tau_{k}\right\}} \chi_{\tau_{k} < t}
\end{align*}
which yields a value between 0 and 1, representing the materiality-weighted defaults
prior to time $t$, $\tau_{k}$ the default time of $GCM_{k}$,
$\theta_{\Xi}$ represents a rate of mean reversion from a stress
state to equilibrium and will be set to 1 in the sequel.

We introduce some thresholds $0 < m_{1} < m_{2} < \dots < m_{S} = 1$ determining the 
stress state, and define the stress state process, $\xi_{t}^{\mathbf{m}}$ taking its value
in $\left\{1,\dots,S\right\}$ by
\begin{align*}
\xi_{t}^{\mathbf{m}} = \left\{ \begin{array}{ll}
1, & \Xi_{t} \leq m_{1} \\
i, & m_{i-1} < \Xi_{t} \leq m_{i}, i \geq 2.
\end{array} \right.
\end{align*}

We introduce $1 = \Lambda_{1} < \dots < \Lambda_{S}$ which 
represent volatility multipliers in each of the $S$ stress states.

We will consider Brownian drivers, $W_{t}^{\xi^{\mathbf{m}}}$,
with regime-dependent volatilities so that
conditional on the stress state $\xi^{\mathbf{m}}$,
the volatility of $W_{t}^{\xi^{\mathbf{m}}}$ is $\xi^{\mathbf{m}}_{t}$:
\begin{align*}
d\left<W^{\xi^{\mathbf{m}}},W^{\xi^{\mathbf{m}}}\right>_{t} = \Lambda_{\xi^{\mathbf{m}}_{t}}^{2} dt.
\end{align*}
Similarly, we will consider regime-switching compound
Poisson driver, $N_{t}^{\xi^{\mathbf{m}}}$,
where, conditional on the value of $\xi^{\mathbf{m}}_{t}$, the intensity of the 
Poisson process is $\lambda \Lambda_{\xi^{\mathbf{m}}_{t}}$, the jump distribution is
as proposed in Inglis et al. \cite{Lipton} being equal in distribution to the random variable
\begin{align*}
e^{Z} - 1
\end{align*}
where $Z \sim N\left(\mu, \sigma\right)$. The Poisson driver is compensated so as to be a martingale
\begin{align*}
\mathbb{E}\left[\left.N_{u}^{\xi^{\mathbf{m}}} - N_{t}^{\xi^{\mathbf{m}}}\right| {\cal{F}}_{t}\right] = 0
\end{align*}
for $u \geq t$.

We remark that the underlying simulation generates the regime-dependent
Wiener and Poisson processes via a numerical scheme based upon superposition
of standard Wiener and Poisson processes over a regime-dependent
number of states. This ensures comparability across different regimes for a particular
realization.

We describe the usage of these drivers below. Note that, losses on default and liquidity drains are primarily driven
by \emph{increments} in the value of portfolios over short time horizons, for this reason, questions related to 
measure-dependent drifts and second-order convexity adjustments are neglected. All processes will be assumed to have a sensitivity to a common regime-dependent
Poisson process that we denote by $N_{t}^{sys, \xi^{\mathbf{m}}}$.

\subsection{Rates Process}
Interest rates in the $i^{th}$ economy are simulated by analogy to a simple 2-factor Hull-White model:
\begin{align}
dr_{t}^{i} =& d\phi_{t}^{i} + dX_{t}^{1} + dX_{t}^{2} + \beta_{i} r_{t-}^{i} dN_{t}^{sys,\xi^{\mathbf{m}}} + r_{t-}^{i} dN_{t}^{i, \xi^{\mathbf{m}}} \label{eq:rates}\\
dX_{t}^{1} =& -\theta_{1} X_{t}^{1} dt + \sigma_{t} dW_{t}^{1,\xi^{\mathbf{m}}} \nonumber \\
dX_{t}^{2} =& -\theta_{2} X_{t}^{2} dt + \alpha \sigma_{t} \left(\rho dW_{t}^{1,\xi^{\mathbf{m}}} + \sqrt{1-\rho^{2}} dW_{t}^{2, \xi^{\mathbf{m}}}\right) \nonumber
\end{align}
where $\phi_{t}^{i}$ is deterministic (used to fit the initial term structure); $N_{t}^{sys,\xi^{\mathbf{m}}},N_{t}^{i,\xi^{\mathbf{m}}}$ are compound, compensated,
regime-dependent Poisson processes of the form considered above representing a systemic (resp. idiosyncratic) jump process;
$\beta_{i}$ represents the sensitivity of rates to the systemic jump process; 
$W_{t}^{1,\xi^{\mathbf{m}}},W_{t}^{2,\xi^{\mathbf{m}}}$ are regime-dependent Wiener Processes of the form considered above which, conditional on
$\xi^{\mathbf{m}}$ are independent;
$\sigma_{t}$ is a deterministic function of time; $\alpha, \theta_{1},\theta_{2},\rho$ are deterministic constants used to control
the relative volatility of rates of different tenors and intra-curve spread volatilities and $t-$ denotes left-hand limit. As described below,
we calibrate the parameters to historical, rather than implied, market data.

Market observables (swap rates, libor rates) are calculated from the state $\left(X_{t}^{1},X_{t}^{2}\right)$
by applying the functional forms derived from the corresponding (affine) two-factor Hull-White model
(without feedback and jump terms). 

The above model may be viewed a minimally complex model that has the key features of:
\begin{itemize}
\item Jumps so as to allow extremal events over short time-periods.
\item Regime-dependent volatilities and intensities so as to capture the natural increasing co-dependence with defaults.
\item Intra-curve spread volatility, so as to ensure a reasonable PL distribution for delta-neutral steepener/flattener positions.
\end{itemize}

\subsection{FX Process}
We model the spot FX analogously, with the spot FX between the $i^{th}$ and $j^{th}$ being governed by
\begin{align}
dX^{i,j}_{t} = \sigma_{t}^{i,j} X^{i,j}_{t} dW_{t}^{i,j,\xi^{\mathbf{m}}} + \beta_{i,j} X^{i,j}_{t-} dN_{t}^{sys, \xi^{\mathbf{m}}}
+ X^{i,j}_{t-} dN_{t}^{i,j,\xi^{\mathbf{m}}} \label{eq:fx}
\end{align}
where $\sigma_{t}^{i,j}$ is a deterministic function of time, $\beta_{i,j}$ is the sensitivity to the systemic jump process
$N_{t}^{sys, \xi^{\mathbf{m}}}$ and $N_{t}^{i,j,\xi^{\mathbf{m}}}$ is an idiosyncratic jump process independent of all else conditional on $\xi^{\mathbf{m}}$.

\subsection{Non-CCP Asset Process}
We model the non-CCP assets of the $k^{th}$ GCM by a process of the same form as (\ref{eq:fx}). 
\begin{align}
dA^{k}_{t} = \sigma_{t}^{k} A^{k}_{t} dW_{t}^{k,\xi^{\mathbf{m}}} + \beta_{k} A^{k}_{t-} dN_{t}^{sys, \xi^{\mathbf{m}}}
+ A^{k}_{t-} dN_{t}^{k,\xi^{\mathbf{m}}} \label{eq:asset}
\end{align}
where each parameter is analogous.

\subsection{Default Events}
The default of each GCM is then determined by a structural model inspired by the Merton-Black-Cox model. More precisely, 
the default time of the $k^{th}$ member is determined as the hitting time of the total position of CCP-related and non-CCP
related activities
\begin{align*}
\tau_{k} = \inf\left\{t>0:C_{t}^{k} - C_{0}^{k} + A_{t}^{k} \leq B_{t}^{k}\right\}
\end{align*}
where $B_{t}^{k}$ is a deterministic barrier; $A_{t}^{k}$ is as above and represents non-CCP assets; $C_{t}^{k} - C_{0}^{k}$
is the net cash-flow due to CCP-related activities (see below).
The barriers are calibrated numerically so as to reproduce target default probabilities.

According to the business model of the member, the contribution to the volatility of a particular member's net assets attributable to
CCP related activity, $C_{t}$, and non-CCP related activity, $A_{t}$, may vary considerably among members. It is useful to categorize members
as follows:
\begin{enumerate}
\item Large diversified financial institutions for which their assets are dominated by non-CCP related activity, $A_{t}$.
\item Large markets-driven houses where the trading book makes up a significant part of their business and for which the volatility 
of $C_{t}$ is substantial relative to that of $A_{t}$.
\item Trading houses for which the volatility of $C_{t}$ may exceed $A_{t}$.
\end{enumerate}

\begin{figure}[htbp]
	\centering
	\includegraphics[trim=0.5cm 0.5cm 0.5cm 0.5cm, clip=true, width=0.8\textwidth]{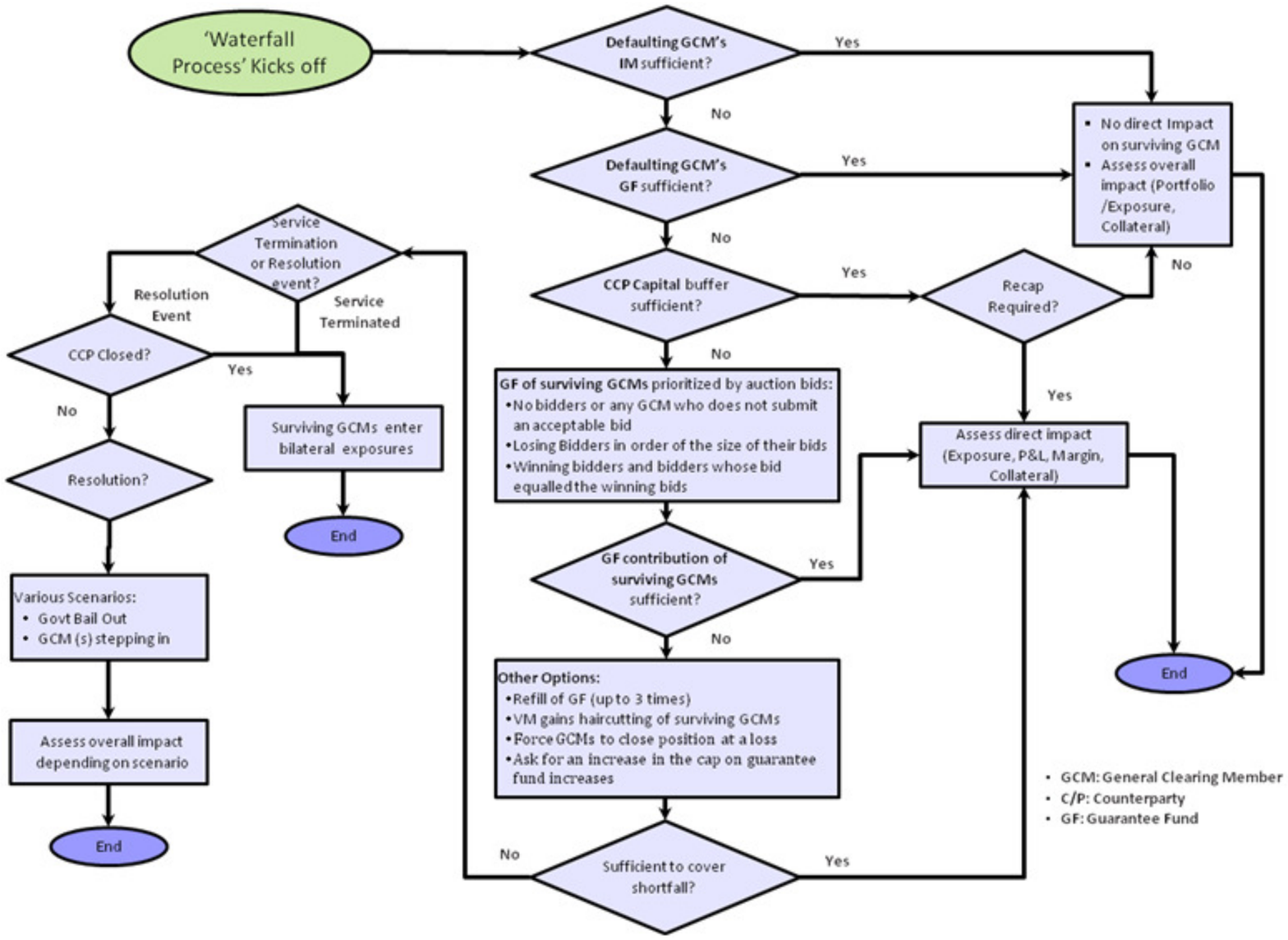}
	\caption{Schematic of CCP Loss Allocation Waterfall}
	\label{fig:defaultWaterfall}
\end{figure}

Once the total loss for a given CCP has been calculated we need to distribute
the losses across the remaining GCMs and go through the standard waterfall
process.
In reality, the method applied will vary considerably between CCPs
and depend on the outcome of the auction process, which is not amenable
to fine modelling. Instead, we
simply distribute the losses among the surviving GCMs in proportion to their IMs,
as though all surviving GCMs bided equally in the auction process.
Similarly, we redistribute the net positions to all surviving members
proportionally to the size of their IM. Figure \ref{fig:defaultWaterfall}, provides a schematic of the loss allocation waterfall.

Finally we need to consider how to model the situation where the losses exceed
all of the CCP's buffers. Namely, that the
losses exceed the total DF and any `end of the waterfall' mitigation
measures the CCP has in place. We make the assumption that surviving GCMs'
will make good the variation on the value of cleared trades up to the time
of default of the CCP and then at this point all trades cleared on the CCP
will be unwound at par. All losses will be divided in the ratio of the
surviving GCM's closing IMs. As there has never been a major CCP default, it is
difficult to say how realistic this resolution is\footnote{One other
possibility is that some combination of the government and the members step in
and take over the running of the CCP however this would add unnecessary
complexity to the model.}. However the authors of this paper believe it to
be a reasonable and parsimonious modelling assumption.

We may summarize this mathematically, as the incremental cash-flows
being represented by
\begin{align*}
C_{t}^{k}-C_{0}^{k} =& -\sum_{CCP_{j}} IM_{CCP_{j}}^{GCM_{k}}\left(t\right) - IM_{CCP_{j}}^{GCM_{k}}\left(0\right)
+ \sum_{CCP_{j}} VM_{CCP_{j}}^{GCM_{k}}\left(t\right) - VM_{CCP_{j}}^{GCM_{k}}\left(0\right) \\
&- \sum_{CCP_{j}} \sum_{t_{i+1} < t} LossIMDF_{CCP_{j}}^{GCM_{k}}\left(t_{i},t_{i+1}\right)
\end{align*}
where the summation is over all CCPs, $IM_{CCP_{j}}^{GCM_{k}}\left(t\right), VM_{CCP_{j}}^{GCM_{k}}\left(t\right)$ are the IM and VM margin,
$LossIMDF_{CCP_{j}}^{GCM_{k}}\left(t_{i},t_{i+1}\right)$ the loss over IM and DF for a $CCP_{j}$ over time
interval $\left(t_{i},t_{i+1}\right]$ allocated to $GCM_{k}$ is given by
\begin{align}
LossIMDF_{CCP_{j}}^{GCM_{k}}\left(t_{i},t_{i+1}\right)
= \frac{\chi_{\tau_{k}>t} IM_{GCM_{k}}^{CCP_{j}} } 
{\sum_{l} \chi_{\tau_{l}>t} IM_{GCM_{l}}^{CCP_{j}}}
LossIMDF_{CCP_{j}}\left(t_{i},t_{i+1}\right)
\end{align}
with the total loss over IM and DF for $CCP_{j}$ given by
\begin{align}
LossIMDF_{CCP_{j}}\left(t_{i},t_{i+1}\right) =
\sum_{l:\tau_{l} \in \left(t_{i},t_{i+1}\right]}
\left(
\sum_{\phi \in \Phi_{CCP_{j}}^{GCM_{l}}\left(t_{i}\right)}\left(V_{\phi}\left(t_{i+1}\right)-V_{\phi}\left(t_{i}\right)\right)
+ IM_{CCP_{j}}^{GCM_{l}}\left(t_{i}\right)
+ DF_{CCP_{j}}^{GCM_{l}}\left(t_{i}\right)
\right)^{-}\label{eq:lossIMDF},
\end{align}
with the summation over those GCM's (if any) that have defaulted in the time interval $\left(t_{i},t_{i+1}\right]$. The length, $t_{i+1}-t_{i}$,
of time intervals in the simulation is aligned to the time horizon corresponding to the VaR methodology of the CCP
(5 business days).

\section{Results}\label{sec:results}
In this section, we present results for a realistic configuration of the model. We take the perspective of a generic
general clearing member `XYZ Bank' proxying one of the `big four' US banks in scale, but with anonymized positions
set to be a fixed proportion of the outstanding gross notional
of each of LCH.SwapClear and CME for USD and EUR fixed-float swaps.
There are a large number of inputs to the model, the setting of which were performed as follows:
\begin{itemize}
\item The parameters describing market dynamics (volatilities, jump intensities, jump distributions and $\beta$'s) were estimated based upon
historic data of suitable representants (eg. 10y swap rate). The systemic jump factor was proxied by a basket of financial equities.
\item GCM and CCP data were based upon publicly available data from the CCP and supplemented by
financial quantities sourced from financial statements of members.
\item We consider a 2-regime configuration of the model, and conservatively\footnote{
The relative realized volatility of EUR rates over the period of the crisis of 2008 to the period 2011-2016 was estimated to be 1.57}
set the value of the stress-regime volatility multiplier, $\Lambda_{2} = 2$, and the boundary between regimes, $m_{1}=0.05$.
\item We suppose there to be 101 clearing members, each members of both of the two CCPs.
\end{itemize}
We wish to consider the distribution at a 1-year time horizon of: (1) the losses due to defaults (of other GCMs and CCPs);
(2) potential liquidity drains on XYZ Bank. These distributions are scaled by the shareholder equity of XYZ bank
since we wish to size the relative significance of losses to capital buffers
and understand the qualitative impact of feedback on the loss distribution.
For the purposes of the example presented here, we set this to 200 BN USD, approximating the magnitude of
the shareholder equity of a `big-four' US bank.

Although we take the perspective of XYZ bank, we reiterate that we take into account
the contingent cash-flows between \emph{all} agents in the CCP network. We study the dependence of these distributions in different
configurations: defaults with the feedback-based regime-switching (labelled `Feedback'); defaults only (labelled `Default Only');
no defaults. We have made use of a minimal entropy path-reweighting algorithm
so as to ensure that the expected stress indicator, $\mathbb{E}\left[\Xi_{1}\right]$, is held fixed as we change settings of
the feedback mechanism so as to ensure comparability of the results across configurations.
The plots are for the complementary cumulative distribution functions
with the y-axis on a logarithmic scale so that, for example, a y-value of 0.01 corresponds to a 99\% quantile of the loss distribution.

\begin{figure}[htbp]
     \centering         
		 \includegraphics[trim=1cm 10cm 1cm 10cm, clip=true, width=1.0\textwidth]{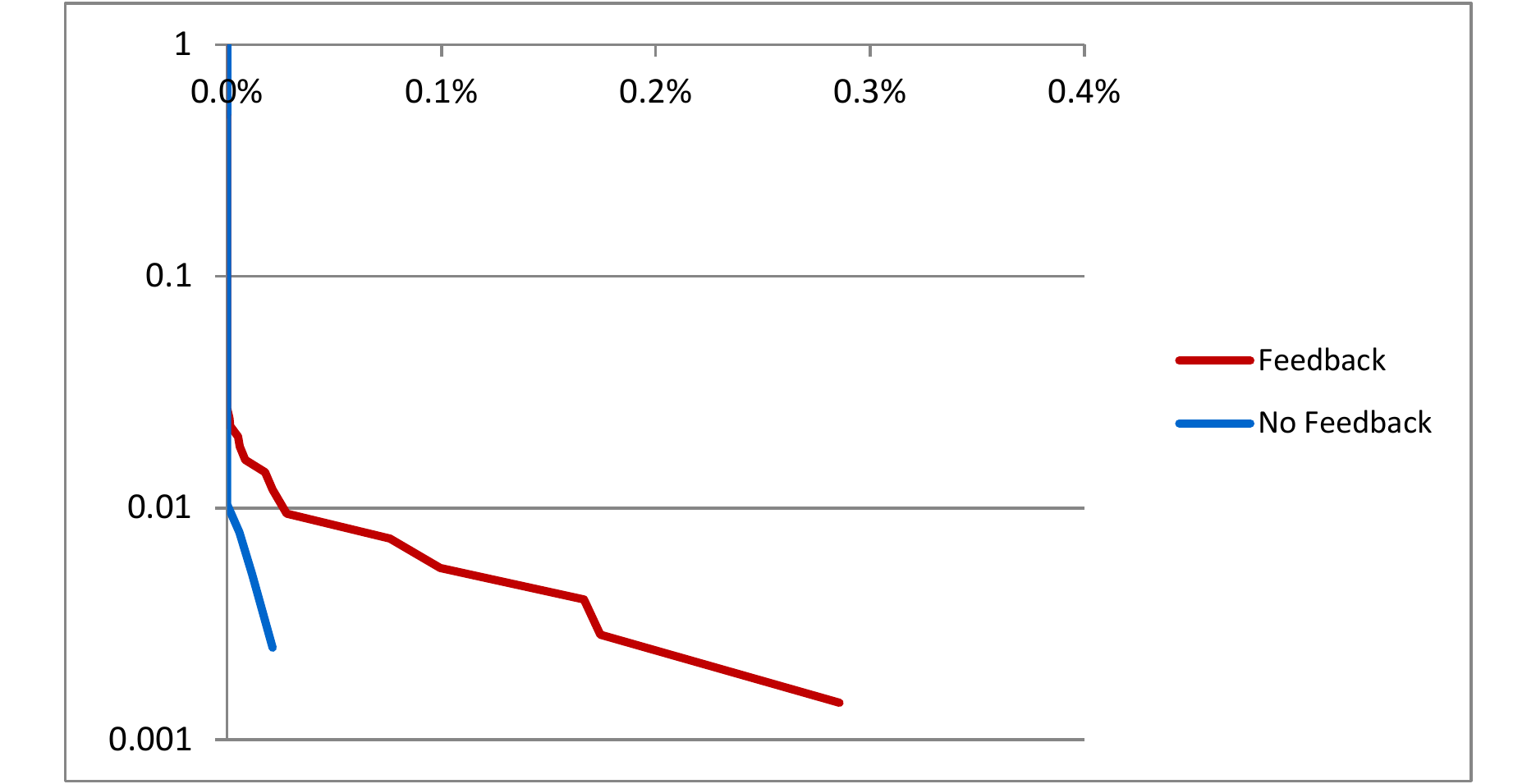}
     \caption{Logarithmic plot of CCDF of Ratio of Losses Due to Default to Equity (\%) of XYZ Bank at time horizon of 1y with and without
	 feedback (logarithmic scale)}
     \label{fig:CDFLosses}
\end{figure}

\begin{figure}[htbp]
     \centering
		 \includegraphics[trim=1cm 9.7cm 1cm 9.7cm, clip=true, width=1.0\textwidth]{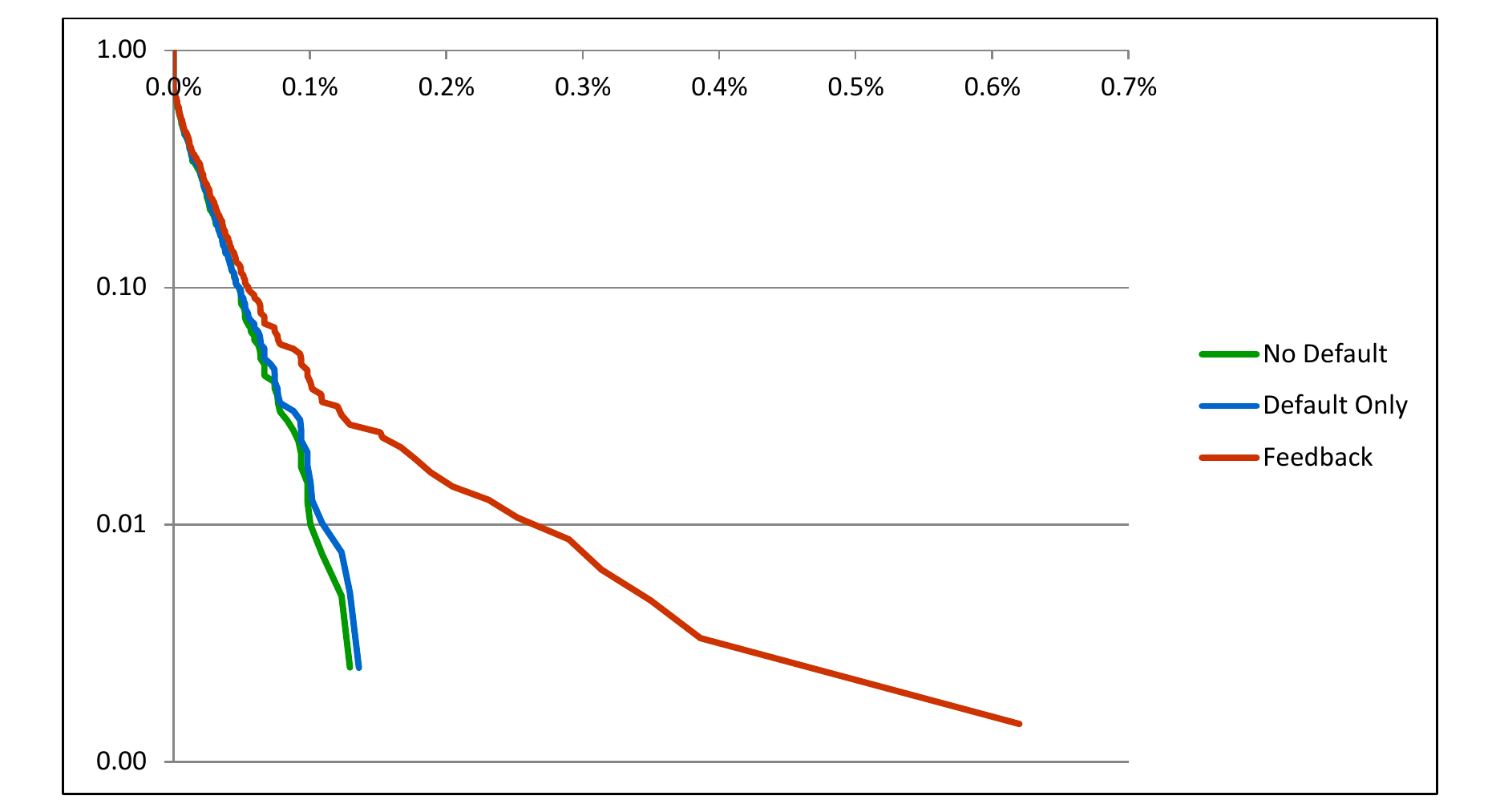}
     \caption{Plot of CCDF of Ratio of Terminal IM-Initial IM to Equity (\%) over both CCPs of XYZ Bank at time horizon of 1y for different model
	 configurations (logarithmic scale)}
     \label{fig:IM}
\end{figure}

Figure \ref{fig:CDFLosses}, presents the simulated distribution of the ratio of the losses due to default
(across all of XYZ's members and CCPs) to its shareholder equity and
demonstrates two key points: firstly, the effect of feedback dramatically amplifies the tail of the
loss distribution due to default, reflecting the importance of capturing the natural wrong-way risk between defaults and market volatility;
secondly, even taking into account the interconnected and complex relationships between the agents of the CCP network,
and making conservative assumptions concerning the relationship between defaults and market volatility, the \emph{scale} of the losses are unlikely to threaten the
survival of a well-diversified and well-capitalized financial institution.

Figure \ref{fig:IM}, presents the simulated distribution for the ratio of the additional aggregate IM to shareholder equity 
required by XYZ Bank posted to the two CCPs in the system. This captures the effect of new extremal events entering the VaR
lookback period and potential increases in portfolio size due to the porting of defaulting members' portfolios\footnote{The
volatility-scaling of the VaR described above is not included.}. The results demonstrate
the significance of capturing the likely increase in volatility in stressed market conditions; further,
it indicates that in dollar-terms liquidity drains due to margining are likely to exceed losses due to default.

\section{Conclusions}\label{sec:conclusions}
Understanding the risks associated to central clearing is technically challenging, since it requires
understanding a large network of GCMs and CCPs and the complex interactions
between them associated to margining, default fund contributions and loss waterfall structures.
We have presented how, using suitable heuristics,
a faithful reflection of the contingent cash-flows between all of the agents in the CCP network may be simulated and the
associated risks investigated. Based upon a model capturing the likely feedback between defaults and volatility,
we have presented results which indicate that the tail losses and
increased liquidity requirements require careful modelling so as to capture the substantial wrong-way risk between volatility
of market variables and defaults, further, liquidity risks dominate those related to credit risk.
Suggesting that, when it comes to members assessing the risks and costs associated to their central clearing activities, 
their primary focus should be on funding and liquidity.

The fear that the wider application of central clearing to OTC derivatives will have a destabilizing impact on the
financial system due to contagion effects transmitted through CCPs are not supported by our experiments.
Primarily, this is attributable to, even conservative bounds on, losses due to default and contingent liquidity requirements being a
small fraction of the tier 1 common equity of the diversified financial institutions that dominate the CCP membership. Although,
the reader should note any CCP-related losses are likely to be realized precisely under the extreme circumstances where
the members are least able to absorb them.

\bigskip

\textbf{Disclaimer:}\emph{ All of the authors are employees of Bank of America.
The views expressed in this paper are those of the authors and do not necessarily represent the views of Bank of America.}

\end{document}